\documentclass[sigconf]{acmart}

\AtBeginDocument{%
  }

\setcopyright{acmlicensed}
\copyrightyear{2024}
\acmYear{2024}
\acmDOI{XXXXXXX.XXXXXXX}

\acmConference[Recsys]{18th ACM Conference on Recommender Systems
}{October 14--18,
  2024}{Bari, Italy}
  
\begin{document}

\title{Item Level Exploration Traffic Allocation in Large-scale Recommendation Systems}


\author{Dong Wang, Junyi Jiao, Arnab Bhadury, Yaping Zhang, Mingyan Gao}
\affiliation{%
  \institution{Google LLC}
  \city{Mountain View}
  \state{California}
  \country{USA}
}
\email{[wangdo, junyijiao, arniebh, yapingzhang, mingyan]  @google.com}
\orcid{0009-0005-9765-6114}

\renewcommand{\shortauthors}{Wang et al.}

\begin{abstract}
This paper contributes to addressing the item cold start problem in large-scale recommender systems, focusing on how to efficiently gain initial visibility for newly ingested content. We propose an  exploration system designed to efficiently allocate impressions to these fresh items. Our approach leverages a learned probabilistic model to predict an item's discoverability, which then informs a scalable and adaptive traffic allocation strategy. This system intelligently distributes exploration budgets, optimizing for the long-term benefit of the recommendation platform. The impact is a demonstrably more efficient cold-start process, leading to a significant increase in the discoverability of new content and ultimately enriching the item corpus available for exploitation, as evidenced by its successful deployment in a large-scale production environment.
\end{abstract}




\received{29 July 2024}
\received[revised]{20 September 2024}
\received[accepted]{20 September 2024}

\maketitle

\section{Introduction}

Large-scale recommendation systems, with millions of new items uploaded daily, face the challenge of effectively recommending fresh content to billions of users due to the lack of initial engagement data. These systems, often based on machine learning, exhibit a feedback loop: models trained on past engagement data tend to favor already popular items, hindering the visibility and accumulation of engagement data for new content. This "item cold-start" problem is a well-known issue, e.g., \cite{ParkAndChu09,Zhang10}.

Exploration, which involves providing initial exposure to new items, is a common approach to address the cold-start problem. By evaluating new items and introducing them to the main recommendation system (exploitation), exploration can help alleviate this issue \cite{Jadidinejad2020}. While exploration might incur a short-term engagement cost due to uncertainty surrounding new items, studies indicate it can expand the pool of recommendable content, benefiting users in the long run \cite{Chen2021}.

Numerous approaches exist to integrate exploration into retrieval or ranking algorithms \cite{Su2024}. In this paper, we focus on a dedicated exploration system that allocates traffic specifically for exploring fresh items. Within this dedicated traffic, new items compete only amongst themselves, not with the general content pool. We address the core problem of optimizing the allocation of billions of exploration impressions to millions of fresh items to maximize a specific positive outcome. This optimization is challenging due to variations in item quality, the need for scalability to handle daily fluctuations in fresh content, and adaptability to changes in the main recommendation system.

We propose a generic approach that aims to solve this problem of traffic allocation for fresh items, ensuring efficiency, scalability, and adaptability in large-scale recommendation systems.

\section{Problem Statement}
We are given a finite pool of $N$ items (indexed $i$ $=$ 1, 2, ..., N) and a total dedicated traffic budget $T$.  The goal is to allocate this traffic across the items, exploring each item $i$ up to a traffic level $X_i$, such that the total allocated traffic does not exceed the budget:
\begin{equation} \label{eq:T}
  \sum_{i=0}^{N} X_i \le T 
\end{equation}. 
The objective is to find the optimal traffic allocation {$X_1$, $X_2$, ..., $X_N$} that maximizes the following cumulative metric:
\begin{equation} \label{eq:G}
  \sum_{i=0}^{N} G_i 
\end{equation},
where $G_i$ is a binary indicator function defined as:
\begin{equation}\label{eq:G_i}
  G_{i} =
    \begin{cases}
      1 & \text{if item is discoverable post exploration}\\
      0 & \text{if item is NOT discoverable post exploration}
    \end{cases}       
\end{equation}.

The desired outcome is defined as "discoverable post exploration". This concept is situated within the framework of exploration and exploitation. An item is classified as discoverable if it can be recommended a minimum number of times by the exploitation engine, signifying that the system has gained sufficient insights into potential user preferences. The exploration phase contributes to an item's discoverability by gathering the initial data required for this understanding.

Furthermore, we assume that exploring each item $i$ incurs a cost $C(X_i)$, and we impose a constraint on the total cost:
\begin{equation} \label{eq:C}
  \sum_{i=0}^{N} C(X_i) \le C_{max} 
\end{equation}.
where $C_{max}$ is the maximum allowable cost for exploration.

\section{Analysis}
We begin by analyzing a simple exploration strategy where each item receives an equal allocation of traffic ($X_i$ $=$ $X_o$ for all $i$), ensuring that $X_o$ $\le$ $T$ / $N$ to satisfy the budget constraint (Equation \ref{eq:T}). However, this approach has several shortcomings:
\begin{itemize}
\item {\textbf{Inefficient exploration of popular items}}: Some items are inherently discoverable ($G_i$ $=$ 1) regardless of exploration. Allocating traffic to these items only increases the cost $C(X_i))$ without any gain in the objective function (Equation \ref{eq:G}).
\item {\textbf{Inefficient exploration of low-quality items}}: Other items may remain undiscoverable ($G_i$ $=$ 0) even after receiving the full allocated traffic $X_o$. This is often the case for low-quality items.  Exploring such items needlessly consumes the exploration budget.
\item {\textbf{Potential under-exploration of promising items}}: There may exist items that would become discoverable if allocated slightly more traffic than $X_o$. These items represent a potentially favorable trade-off between the gain in discoverability and the associated cost.
\end{itemize}

These observations highlight the potential for optimization by assigning different traffic levels based on item-specific characteristics. This leads to the following key question:

\textit{Given a set of features describing item $i$, can we predict the minimum traffic level $X_i$ required to make it discoverable?}

\section{Modeling}
\subsection{Formulation}
To predict the minimum traffic needed for item $i$ to become discoverable, we propose a modeling approach based on experimental data collection. In this process, we explore each item up to a traffic level $Y_i$ and observe the resulting discoverability ($G_i$). However, the challenge lies in the fact that $Y_i$ does not necessarily equal $X_i$. In fact, any $Y_i$ $\ge$ $X_i$ could result in $G_i$ $=$ 1.

To address this, we introduce an intermediate modeling problem that aligns better with our data collection process:

\textit{Given the features of item $i$ and a traffic level $Y_i$, what is the probability that the item will become discoverable (i.e., $P(G_i = 1 | Y_i)$)?}

This is formulated as a classification problem. Once we have learned $P(G_i = 1 | Y_i)$, we can determine $X_i$ by setting a desired confidence level $CF$ and solving the following equation:

\begin{equation} \label{eq:dynamic_cap}
P(G_i = 1 | X_i) = CF.
\end{equation}
Here, $CF$ represents the probability of achieving discoverability when allocating $X_i$ traffic to item $i$.

\subsection{Model Architecture}
\begin{figure}[h]
  \centering
  \includegraphics[width=\linewidth]{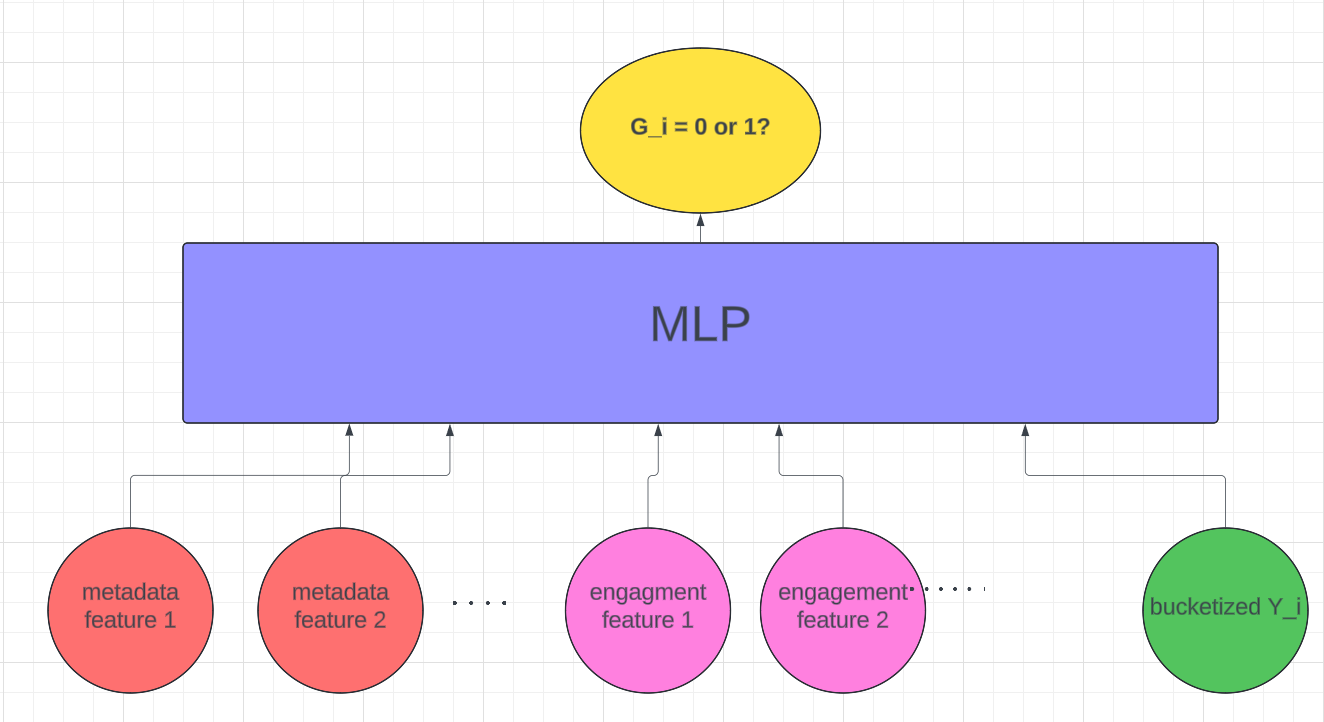}
  \caption{Model architecture for predicting $P(G_i = 1 | Y_i)$.}
  \Description{$Y_i$ can be any integer, but for modeling purpose here we discretize $Y_i$. The probability condition
  $Y_i$ bucket is fed into the model as a discrete feature.}
\end{figure}

To model $P(G_i = 1 | Y_i)$, we employ a standard classification approach with cross-entropy loss. The model utilizes various item metadata that could be predictive of discoverability. Furthermore, we leverage engagement data collected during exploration to enhance model performance.

Recognizing the influence of $Y_i$ on $P(G_i = 1 | Y_i)$, we discretize $Y_i$ into buckets and experiment with different ways of incorporating it into the model:

\begin{itemize}
\item {Direct concatenation with other item features.}
\item{Concatenation at intermediate layers within the model.}
\item{Attention mechanisms between $Y_i$ and other features.}.
\end{itemize}

We acknowledge that during serving time, features may become available at different stages, and we can trigger multiple inferences to refine the prediction.

\subsection{Example Output}
\begin{figure}[h]
  \centering
  \includegraphics[width=\linewidth]{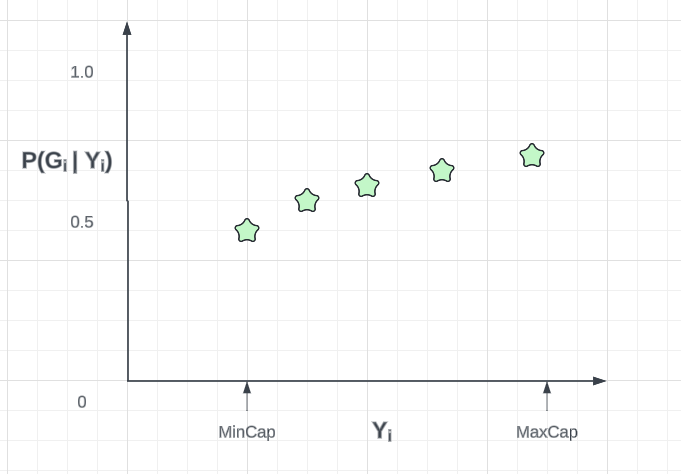}
  \caption{An example model output for an item i}
  \label{fig:example_output}
  \Description{The $P(G_i | Y_i)$ is predicted for each bucket of $Y_i$.}
\end{figure}

Figure \ref{fig:example_output} illustrates an example model output, depicting the predicted $P(G_i = 1 | Y_i)$ for each bucket of $Y_i$. Additionally, we introduce two constants, $MinCap$ and $MaxCap$, to define a parameter space for $X_i$:
\begin{equation}
MinCap \le X_i \le MaxCap.
\end{equation}

\section{Traffic Allocation}\label{section:allocation}
The model developed in the previous section provides predictions for $P(G_i = 1 | Y_i)$, the probability of an item becoming discoverable given a certain traffic level. However, this does not directly specify the optimal traffic allocation. To address this, we present a scalable and adaptable traffic allocation approach.

\begin{figure}[h] 
  \centering
  \includegraphics[width=\linewidth]{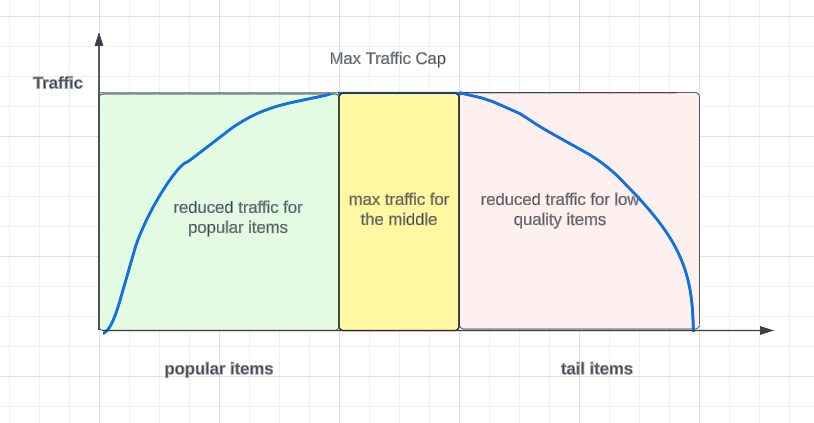}
  \caption{Item Level Traffic Allocation Approach}
  \label{fig:traffic_allocation}
  \Description{For efficient use of the exploration traffic, we designed a strategy to allocation exploration views that are customed to each item's properties.}
\end{figure}

Our strategy involves partitioning the items into three regions based on their predicted discoverability at $MaxCap$:

\begin{itemize}
\item {\textbf{High-Discoverability Region, $P(G_i = 1 | MaxCap)$ $>$ $CF_{high}$}}: 
Items in this region are likely to become discoverable with less traffic than $MaxCap$. To efficiently allocate resources, we determine $X_i$ by solving Equation \ref{eq:dynamic_cap} with a confidence level of $CF_{high}$.
\item{\textbf{Moderate-Discoverability Region, $P(G_i = 1| MaxCap)$ $\ge$ $CF_{low}$ $\&\&$ $P(G_i | MaxCap)$ $<$ $CF_{high}$}}: 
Items in this region require the maximum exploration effort to achieve discoverability. We allocate the full $MaxCap$ traffic to these items.
\item{\textbf{Low-Discoverability Region, $P(G_i = 1| MaxCap)$ $<$ $CF_{low}$}}: 
Items in this region have a low probability of becoming discoverable even with $MaxCap$ traffic. While allocating the full $MaxCap$ might be inefficient, some exploration is still warranted to account for potential model errors and to ensure the exploration algorithm doesn't become overly exploitative. Therefore, we allocate traffic to these items proportionally to their user feedback, with better feedback leading to higher (but still capped at $MaxCap$) exploration traffic.   
\end{itemize}

Large recommendation systems are constantly evolving. To ensure our approach's effectiveness in such dynamic environments, we propose two key adaptations:

\begin{itemize}
\item{\textbf{Adjusting Traffic Allocation based on Item/Traffic Growth}} As the system grows, the number of items and total exploration traffic may change at different rates. We can adapt the traffic allocation within the "Low-Discoverability Region" to maintain the efficiency of discoverable corpus generation. Specifically, we reduce allocation to this region if item growth outpaces traffic growth, and vice versa.
\item{\textbf{Continuous Model Retraining}}:
 The evolving nature of the recommender system may change the exploration traffic requirements for items. We address this by continuously retraining the $P(G_i = 1 | Y_i)$ model on the latest production data, ensuring it remains accurate and reflective of current user behavior.
\end{itemize}

\section{Results}
The model demonstrated strong predictive performance, achieving an overall Area Under the Receiver Operating Characteristic Curve (AUC) score of 0.96 and a Precision-Recall AUC (PR-AUC) score of 0.84.  Performance remained consistent across all traffic buckets ($Y_i$). Further model performance metrics are detailed below.

\begin{figure}[h]
  \centering
  \includegraphics[width=\linewidth]{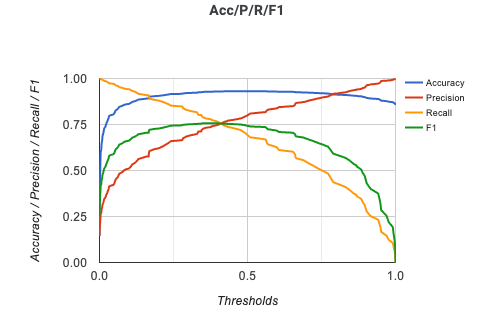}
  \caption{Accuracy, Precision, Recall, and F1 for the model predicting whether an item is discoverable.}
  \label{fig:model_metric_1}
  \Description{Accuracy, Precision, Recall, and F1 for the model predicting whether an item is discoverable.}
\end{figure}

\begin{figure}[h]
  \centering
  \includegraphics[width=\linewidth]{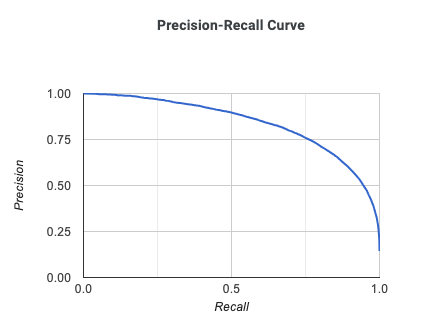}
  \caption{Precision-Recall Curve for the model predicting whether an item is discoverable.}
  \label{fig:model_metric_2}
  \Description{Precision-Recall Curve for the model predicting whether an item is discoverable.}
\end{figure}

We find that incorporating engagement rate features significantly enhances model performance. To leverage the most recent engagement data during online serving, a multi-inference approach was implemented. This allows for updated engagement features to be utilized at various points throughout an item's lifecycle, further optimizing model performance.

We implemented the traffic allocation algorithm as described in section \ref{section:allocation}. We observed the desired shift of traffic from high-popularity items to those in the tail distribution. The enhanced efficiency in traffic allocation yielded a measurable impact on key performance indicators. This system has been successfully deployed to production and is currently serving users.

\bibliographystyle{ACM-Reference-Format}
\bibliography{sample-base}

\end{document}